**Electroencephalography and mild cognitive impairment research: A scoping review and bibliometric analysis (ScoRBA)**

**Short running title:** EEG and mild cognitive impairment

Adi Wijaya*, Noor Akhmad Setiawan†, Asma Hayati Ahmad‡, Rahimah Zakaria‡, Zahiruddin Othman¶

*Department of Health Information Management, Universitas Indonesia Maju, Jakarta, Indonesia, , †Department of Electrical and Information Engineering, Universitas Gadjah Mada, Yogyakarta, Indonesia, ‡Department of Physiology, School of Medical Sciences, Universiti Sains Malaysia, Kubang Kerian 16150, Kelantan Malaysia, , ¶Department of Psychiatry, School of Medical Sciences, Universiti Sains Malaysia, Kubang Kerian 16150, Kelantan, Malaysia

**Correspondence:** Zahiruddin Othman, Department of Psychiatry, School of Medical Sciences, Universiti Sains Malaysia, Kubang Kerian 16150, Kelantan, Malaysia. E-mail: zahirkb@usm.my

**Conflict of interest**

The authors declare no conflict of interest.

**Source of funding**

This research did not receive any specific funding.

## ABSTRACT

*Background:* Mild cognitive impairment (MCI) is often considered a precursor to Alzheimer's disease (AD) due to the high rate of progression from MCI to AD. Sensitive neural biomarkers may provide a tool for an accurate MCI diagnosis, enabling earlier and perhaps more effective treatment. Despite the availability of numerous neuroscience techniques, electroencephalography (EEG) is the most popular and frequently used tool among researchers due to its low cost and superior temporal resolution.

*Objective:* We conducted a scoping review of EEG and MCI between 2012 and 2022 to track the progression of research in this field.

*Methods*: In contrast to previous scoping reviews, the data charting was aided by co-occurrence analysis using VOSviewer, while data reporting adopted a Patterns, Advances, Gaps, Evidence of Practice, and Research Recommendations (PAGER) framework to increase the quality of the results.

*Results:* Event-related potentials (ERPs) and EEG, epilepsy, quantitative EEG (QEEG), and EEG-based machine learning were the research themes addressed by 2310 peer-reviewed articles on EEG and MCI.

*Conclusion:* Our review identified the main research themes in EEG and MCI with high-accuracy detection of seizure and MCI performed using ERP/EEG, QEEG and EEG-based machine learning frameworks.

**Key messages:**

- Health librarians and researchers may conduct a data charting for scoping review using co-occurrence analysis.
- Health librarians may engage various review methods to promote more research and publications related to health information services.
- Researchers are encouraged to conduct further research based on the research themes identified in this study.

## Introduction

The most common type of neurodegenerative dementia is Alzheimer's disease (AD), with ten million new dementia cases identified each year (World Health Organization [WHO], 2020). AD is the most common cause of dementia accounting for 60–70% of these new cases, followed by vascular dementia (VaD), dementia with Lewy bodies (DLB), and other forms of neurodegenerative illnesses (WHO, 2020).

Prior to development of dementia, two stages of decreased cognition are often defined: subjective cognitive decline (SCD), which is not corroborated by an informant or neuropsychological testing, and mild cognitive impairment (MCI), which is corroborated by an informant or neuropsychological testing (Albert et al., 2011; American Psychiatric Association. Diagnostic and Statistical Manual of Mental Disorders [DSM], 2013; Jessen et al., 2014; Winblad et al., 2004; WHO, 1992). In both stages, individuals are able to undertake personal and instrumental daily activities (Gao et al., 2018; Mauri, Sinforiani, Zucchella, Cuzzoni, & Bono, 2012; Pandya, Lacritz, Weiner, Deschner, & Woon, 2017; Visser, Kester, Jolles, & Verhey, 2006). According to the Alzheimer Association (2020), 15% of MCI in individuals over 65 progress to dementia, but in a different study (Chen et al., 2020), 32% of individuals with MCI were found to have an AD at the 5-year follow-up. Older age, poor cognition, APOE 4 allele carrier status, and hypertension could enhance dementia risk (Roberts & Knopman, 2013). Thus, it is important to establish biomarkers that can identify those at high risk for dementia and, ideally, its cause.

Studies combining cognitive assessments with cerebrospinal fluid (CSF) proteins, Magnetic Resonance Imaging (MRI), and Fluorodeoxyglucose-Positron Emission Tomography (FDG-PET) have shown outstanding results in detecting MCI patients who will later develop dementia (Ottoy et al., 2019; Smailagic, Lafortune, Kelly, Hyde, & Brayne, 2018; Zhou et al., 2019). However, these biomarkers are time-consuming, expensive, and some are invasive, therefore not suitable for daily clinical practice, except in subspecialized settings. Electroencephalography (EEG) is a low-cost, noninvasive, and straightforward method that can be performed in most clinics and even in primary care settings, especially with automatic reading. In the recent decade, many studies have published the EEG's ability to diagnose MCI patients who will develop dementia (Bonanni et al., 2015; Cassani, Estarellas, San-Martin, Fraga, & Falk, 2018; Moretti, 2015a&b; Poil et al., 2013). Therefore, this study aims to perform a scoping review (Arksey & O'Malley, 2005) aided by bibliometric analysis (ScoRBA) on EEG and MCI research, and utilise the Patterns, Advances, Gaps, Evidence of Practice, and Research

Recommendations (PAGER) framework (Bradbury-Jones, Isham, Morris, & Taylor, 2019) to present our data.

## Methods

### Literature search and study selection

The scoping review's five steps were adapted from Arksey and O'Malley's (2005) methodology (Table 1). Step 1, which was to identify the primary research question, was deliberated in the introductory section. On September 20, 2022, a literature search using the Scopus database was carried out (step 2). We decided to use the Scopus database because it covers a large number of publications (Pranckute, 2021; Zhu & Liu, 2020). The following search phrases: "TITLE-ABS-KEY (EEG OR electroencephalogra*)" AND TITLE-ABS-KEY ("neurocognitive disorder" or "cognitive impairment" or "cognitive disorder" or "cognitive disability" OR MCI) were used. We only included journal articles that were published in the English language between 2012 and 2022 (step 3). The PRISMA flow diagram in Figure 1 exhibits the search procedure.

**Table 1** Five-step scoping review framework

| Step | Description |
|---|---|
| 1. Identification of Research Questions and Related Studies | The research question in this paper is “What are the main themes in the EEG and mild cognitive impairment research?” |
| 2. Identification of Related Studies | A preliminary reading of relevant studies was performed to determine the keywords and terms utilised in the article selection |
| 3. Study Selection | Peer-reviewed original articles published in the Scopus-indexed journal that were written in English were selected. Articles that assessed the usage of EEG in patients with MCI met the selection criteria of this study. |
| 4. Charting the data | The co-occurrence keyword was used to group the data into a graphical representation. |
| 5. Collating, summarizing, and reporting the results | The Patterns, Advances, Gaps, Evidence of Practice, and Research Recommendations (PAGER) framework was used to summarize, analyze, and present the results in order to enhance the quality and applicability of the literature review, as suggested by Bradbury et al. (2019) |

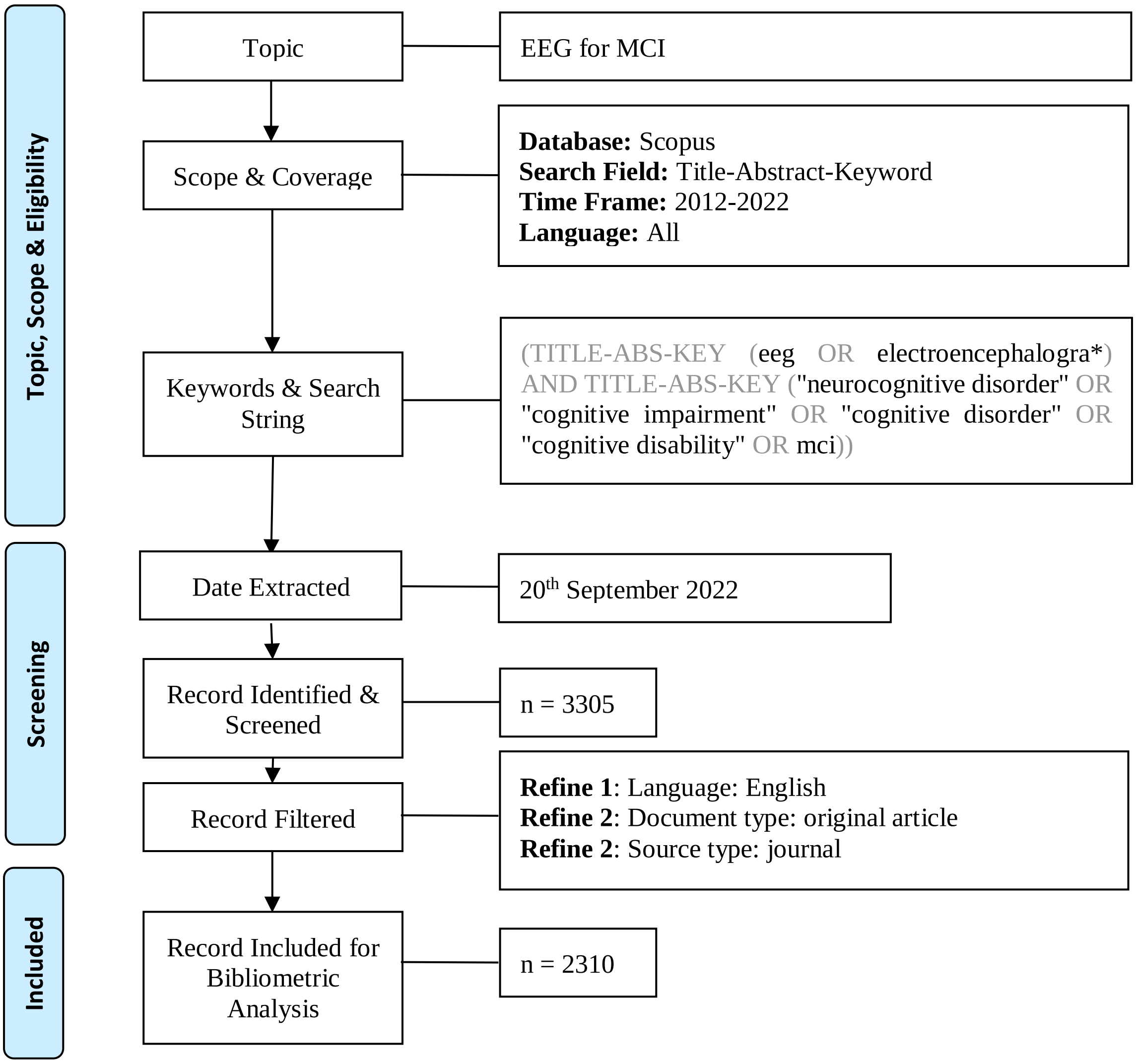

**Figure 1** PRISMA flow diagram of the search strategy (Page et al., 2021)

## Results

The initial search results from the Scopus database generated 3305 documents published from 2012 until September 2022 (Figure 1). After removing papers in languages other than English, document types other than original articles (primary research documents), and source types other than journals, 2310 documents remained.

Publication output

The number of articles showed an increasing trend from 2012 to 2021. There was a drop in the number of articles in 2022 due to incomplete data (Figure 2).

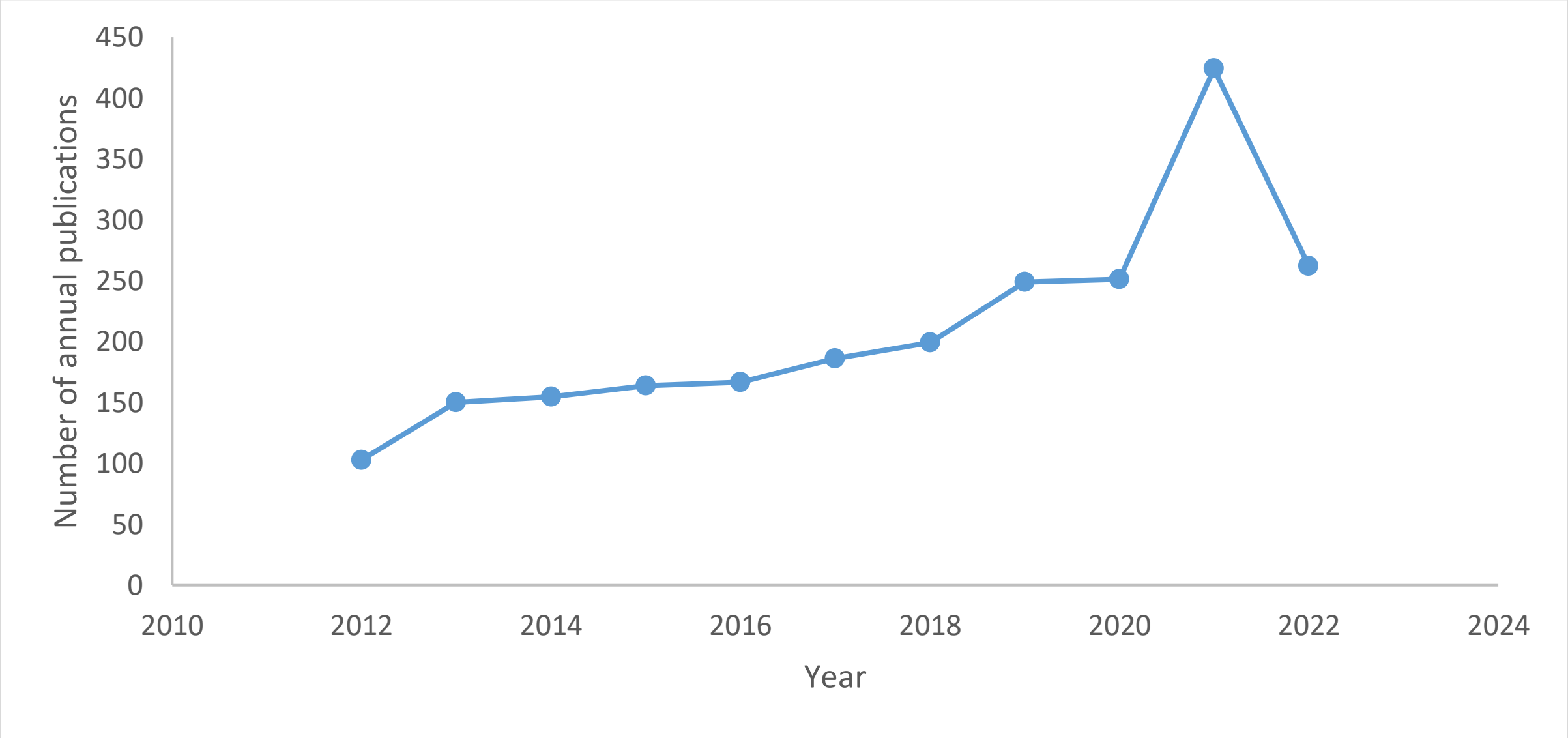

**Figure 2** The number of annual publications on EEG and MCI research published from 2012 to September 2022 in the Scopus database

Research output on EEG and MCI research covered 24 different subject areas. Medicine, neuroscience, biochemistry, genetics and molecular biology, and psychology were the four subject areas that made up the most articles (3521/4172, or 84.40% of the total). These were followed by computer science, engineering, pharmacology, toxicology and pharmaceutics, multidisciplinary, nursing, health professions, etc. Figure 3 lists the top 20 EEG and MCI research areas currently under investigation.

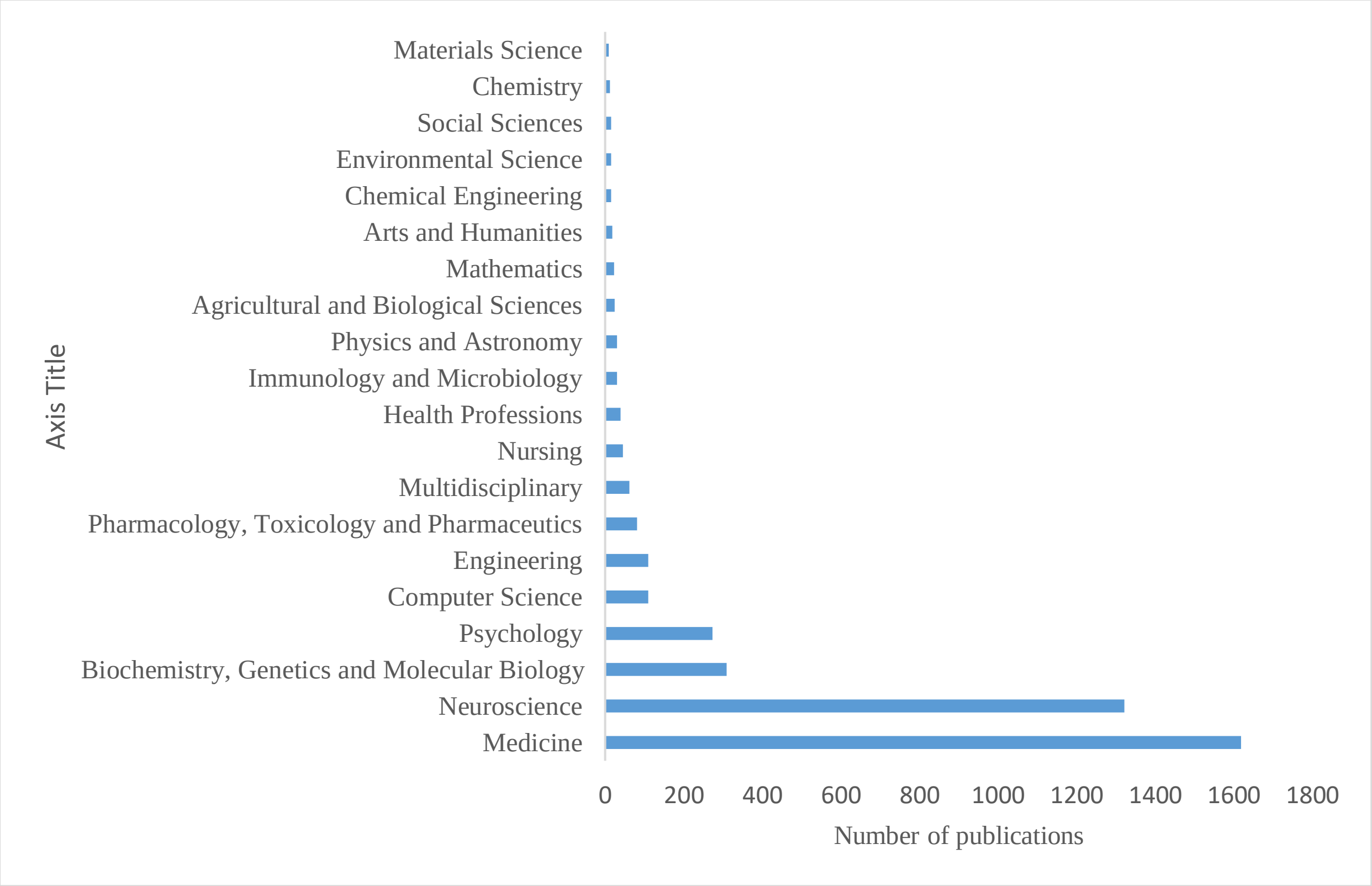

**Figure 3** Top 20 subjects in the area of EEG and MCI research

**Data Charting**

Previous method of data charting using the 'narrative review' approach (Pawson, 2002) was replaced with utilizing the results from the co-occurrence keyword analysis. All the 2310 documents were downloaded from the Scopus database in Comma-Separated Values (.csv) format and analysed using VOSviewer version 1.6.17 (van Eck & Waltman, 2021) for co-occurrence keyword analysis. A total of 5002 author keywords were collected from our collection of 2310 publications. Only keywords that appeared at least five times were selected for keyword analysis. This criterion was met by 229 keywords out of 5002. Our analysis indicates that the keyword network has four distinct clusters as shown in Figure 4: cluster one (red, 80 keywords), cluster two (green, 56 keywords), cluster three (blue, 47 keywords) and cluster four (yellow, 46 keywords).

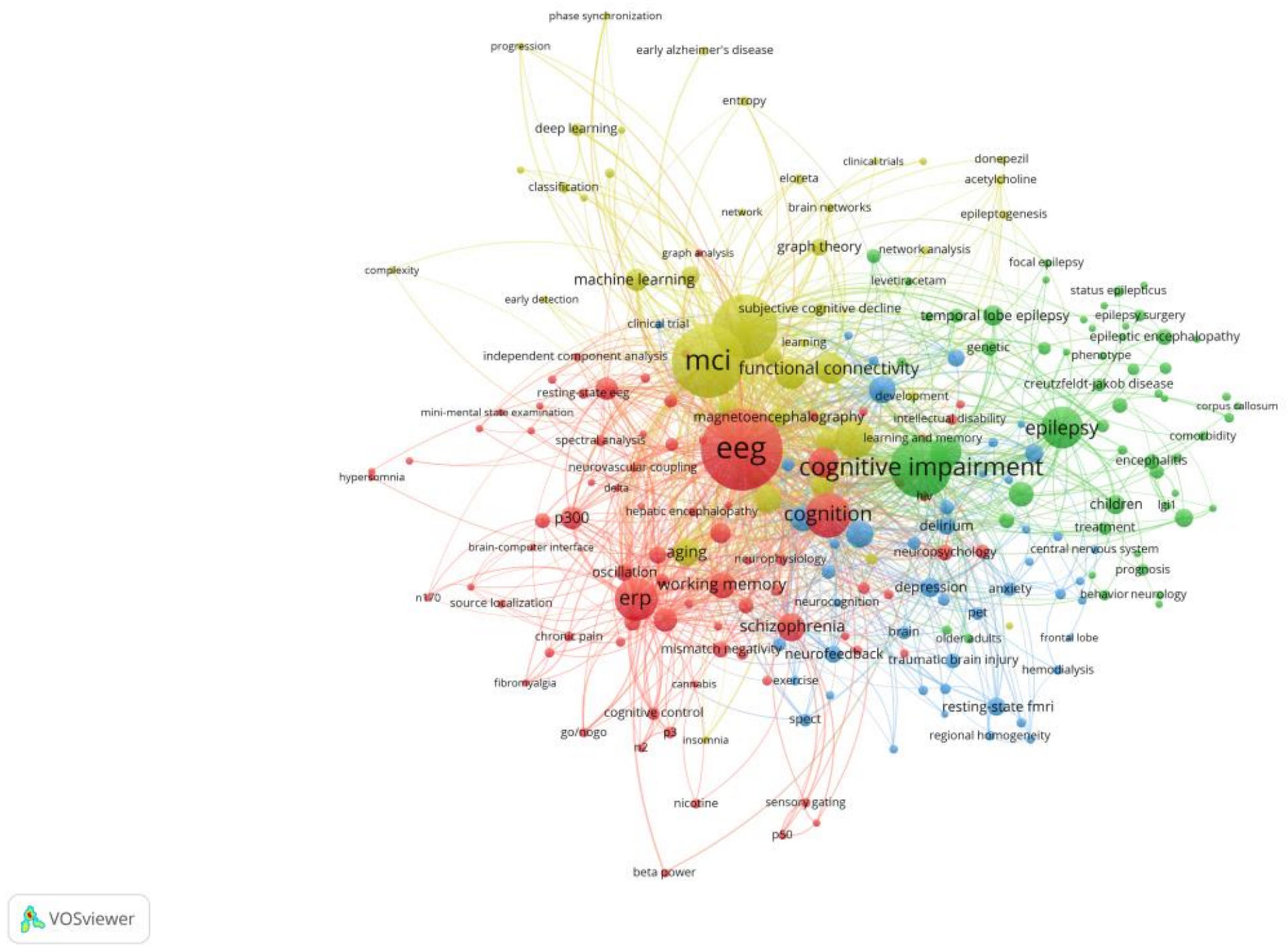

**Figure 4** Co-occurrence author keywords. 229 out of 5002 keywords met the threshold of a minimum five occurrence

Based on the dominant keywords in each cluster, research on the usage of EEG in mild cognitive impairment patients can be divided into four broad clusters and themes: (1) red cluster – event-related potentials (ERPs) and EEG, (2) green cluster - epilepsy, (3) blue cluster – quantitative EEG, and (4) yellow cluster – EEG and machine learning.

## DISCUSSION

The PAGER framework was utilised as the final step of our scoping assessment of the usage of EEG in mild cognitive impairment research (Bradbury-Jones et al., 2019). Table 2 summarizes the key findings of the PAGER framework literature analysis, including model themes, research advances, research gaps, evidence of practice, and research recommendations.

**Table 2** Results of the PAGER analysis of the EEG and mild cognitive impairment research

| Patterns | Advances | Gaps | Evidence for practice | Research recommendation |
|---|---|---|---|---|
| ERP/EEG (Red cluster) | Discussion on the combined EEG/ERP measures of memory as a diagnostic tool for detecting MCI or prodromal AD. | There is a need to understand the causal effects between Aβ depositions and neural excitability. | Evidence from future research on EEG/ERP as biomarkers for assessing early memory decline and treatment response. | Future studies to establish ERPs' sensitivity and specificity in discriminating AD, MCI and cognitively intact AD-risk groups as well as measuring late cognitive ERPs during complicated tasks, especially in healthy older adults at risk for cognitive decline. |
| Epilepsy (Green cluster) | Knowledge discovery from machine learning classifiers such as seizure localization, which exactly points to affected brain lobe(s), channel importance, and based on participating channels in a seizure may help to predict cognitive impairment | There is a need to select machine learning classifiers and features for high-accuracy seizure detection and predict cognitive impairment. | EEG is currently used to detect seizures and categorise seizure types and epilepsy disorders. | Future studies to explore the associations between cognitive deficit and seizure timing. |
| | secondary to seizures, EEG interictal spikes (IIS), and antiepileptic medicines. | | | |
| QEEG and neuropsychiatry (Blue cluster) | Development of simple diagnostic algorithms using spectral analysis of EEG to detect MCI cases. | There is a need to find clear-cut EEG signs and correlations with neuropathology and neuropsychology for consistent and precise assessments. | Evidence from future research on QEEG as a diagnostic and prognostic biomarker as well as for monitoring treatment response in MCI cases. | Future large-scale longitudinal clinical investigations are required to determine the diagnostic and prognostic potential of QEEG measurements as early functional markers of AD on a subject-by-subject basis. |
| EEG and machine learning (Yellow cluster) | Machine learning for extracting the relevant EEG/ERP data for cognitive assessment. | There is a need to select machine learning classifiers and features for a high-accuracy cognitive assessment. | Evidence from future research to apply EEG-based machine learning frameworks for accurate diagnosis of MCI cases. | Future studies to utilize an EEG-based machine learning framework for extracting the most relevant features from EEG data empirically thereby resulting in a high accuracy of cognitive assessment. |

**EEG/ERPs and mild cognitive impairment**

The first report on using scalp EEG was in 1929 (Berger, 1929). EEG is a cost-effective and non-invasive technology that directly assesses the mean electrical activity of the brain at scalp sites with excellent temporal precision (milliseconds) (Cohen, 2017). ERPs are one of the most extensively used EEG techniques to examine brain activity in response to sensory, motor, or cognitive events. Several ERP components (e.g., N1, P2, and P3) can provide full information about low or high level cognitive activities of the brain (Luck, 2014) and have been used in research for several decades.

Three EEG measures that have been found to strongly correlate with verbal learning and memory abilities in healthy elderly people and patients with MCI or prodromal Alzheimer's disease (AD) in previous research include ERP P600 (Olichney et al., 2002, 2006 & 2013), suppression of oscillatory activity in the alpha frequency range (Fell, Ludowig, Rosburg, Axmacher, & Elger, 2008; Fellner, Bauml, & Hanslmayr, 2013; Hanslmayr, Spitzer, & Bauml, 2009; Sederberg et al., 2006; Vassileiou, Meyer, Beese, & Friederici, 2018), and cross-frequency coupling between low theta/high delta and alpha/beta activity. A study by Xia and colleagues (2020) evaluated the relationships among three previously identified electrophysiological measures: the P600 ERP and two oscillatory effects (i.e. alpha suppression and ϕ/δ-α/β coupling), and their potential as biomarkers of verbal memory functioning in healthy ageing, MCI and prodromal Alzheimer’s disease. The study highlighted the importance of combining ERP and EEG oscillatory measures as predictors of verbal memory in MCI individuals. Additionally, a recent study by Meghdadi et al. (2021) demonstrated that subtle resting state EEG abnormalities could differentiate between MCI and AD groups. The study designed classifiers for differentiating individual MCI and AD participants from age-matched controls and concluded that the MCI group demonstrated a moderate increase in δ, ϕ, and Theta-to-Alpha (TAR) ratio, mostly in temporal channels (Meghdadi et al., 2021).

A systematic literature review investigating ERPs as potential biomarkers of AD-related neuropathology was published recently (Paitel, Samii, & Nielson, 2021). The patterns in cognitive ERPs (≥150 ms post-stimulus) differentiate MCI, AD, and cognitively intact elders who carry AD risk through the Apolipoprotein-E ε4 allele (ε4+), from healthy older adult controls. The review also implies that the integration of ERPs into cognitive assessment may significantly improve early identification and characterization of brain dysfunction, allowing for faster differential diagnosis and intervention targeting (Paitel et al., 2021). Future studies are required to establish ERPs' sensitivity and specificity in discriminating AD, MCI and cognitively intact AD-risk groups as well as measuring late cognitive ERPs during complicated tasks, especially in healthy older adults at risk for cognitive decline.

In a more recent study, Devos et al. (2022) found that neuronal hyperexcitability is related to higher amyloid levels in cognitively healthy older persons. The smaller P3 ERP task difference suggests a less efficient stimulus processing in cognitively healthy older persons with elevated amyloid levels compared to those with non-elevated amyloid levels. These results, together with the lack of statistically significant differences in behavioural outcomes, imply that the enhanced Aβ deposition-induced hyperexcitability is a non-functional neuronal compensation that may later impair working memory. There is a need to understand the causal effects of Aβ depositions and neural excitability to confirm this observation.

**Epilepsy and mild cognitive impairment**

Epileptic seizures are brief bursts of excessively synchronised brain activity (Chang & Lowenstein, 2003; Fisher, Scharfman, & DeCurtis, 2014; Noachtar et al., 1999; Shorvon, 2011; Trevelyan, 2016; Wang et al., 2017). They cover a wide range of occurrences, from mild clinical manifestations, such as brief and hardly detectable loss of consciousness, to vigorous episodes of muscular shaking that may cause physical harm (Landi, Petrucco, Sicca, & Ratto, 2019). Epilepsy is diagnosed based on EEG abnormalities, including seizures and interictal epileptiform discharges (IEDs). EEG findings help to categorise seizure types and epilepsy disorders (Chen & Koubeissi, 2019).

EEG analysis of patients with recurrent seizures shows "subclinical" aberrant electrical activity between seizures. IEDs are classified based on their EEG signature: sharp waves, spikes, sharp waves/spikes-and-slow waves, and numerous spikes-and-slow waves (de Curtis & Avanzini, 2001; Fisher et al., 2014; Gotman, 1980; Kane et al., 2017; Kooi, 1966; Noachtar et al., 1999; Pillai & Sperling, 2006). However, background noise and aberrations such as eye blinks and muscle movements can contaminate EEG readings, causing electrical interference that is difficult to detect visually in longer recordings. Thus, there is a need to select machine learning classifiers and features for high-accuracy seizure detection and predict cognitive impairment (Natu, Bachute, Gite, Kotecha, & Vidyarthi, 2022).

Although seizures are the most obvious clinical manifestation of epilepsies, people with epilepsy are at risk of other health problems at a higher incidence rate than would be expected by chance (Institute of Medicine [IOM], 2012). Common co-morbidities in epilepsy include cognitive impairment such as memory, attention, and processing problems, mental health problems such as depression and anxiety, and somatic co-morbidities such as sleep disorders and migraines (Holmes, 2015).

Most of the cognitive impairment in epilepsy is due to its aetiology. Trauma, hypoxia, ischemia, and mesial temporal sclerosis secondary to prolonged seizures are acquired illnesses that can cause epilepsy and cognitive impairment. Genetic disorders including tuberous sclerosis, Fragile X, Rett, and Dravet syndromes can also lead to epilepsy and cognitive impairment. In addition to static abnormalities induced by the underlying aetiology, cognitive impairments might be temporary due to seizures, EEG interictal spikes (IIS), and antiepileptic medicines. In many people, cognitive impairment is due to a combination of causes (Holmes, 2015).

The age of seizure onset is another factor that is associated with cognitive impairment. When compared to healthy peers, children with epilepsy are more likely to have severe cognitive impairments, which primarily affect language, semantic, motor, and visual functions (Al-Malt, Abo Hammar, Rashed, & Ragab, 2020). According to the same study, children with idiopathic epilepsy who have more nocturnal seizures than diurnal seizures have both overt and subtle cognitive abnormalities (Malt et al., 2020). Future studies are needed to explore the associations between cognitive deficit and seizure timing.

**Quantitative EEG and mild cognitive impairment**

The diagnosis of cognitive impairment and neurodegenerative illnesses using EEG relies heavily on the EEG expert's visual assessment. The limitations that come with this have been supplemented by methods such as quantitative EEG (QEEG) that extracts particular parameters from continuous variable data analysis. At current, studies on brain function assessment and early diagnosis of degenerative brain illnesses such as AD and AD-MCI are actively advancing (Al-Qazzaz et al., 2014; Cassani et al., 2018; Smailovic & Jelic, 2019). For instance, QEEG measurements appear to differentiate between different forms of dementia and correlate favourably with surrogate markers of AD neuropathology, making them promising low-cost and noninvasive diagnostics of AD (Smailovic & Jelic, 2018). The diagnostic and prognostic value of QEEG measurements as early functional markers of AD should be determined in future extensive longitudinal clinical studies on a subject-by-subject basis.

One technique for analysing QEEG signals is using the power spectrum, and this approach is a well-established core component of the general method utilised in clinical research. The most frequent spectral shift so far is thought to be connected to the degree of cognitive deterioration brought on by the rise in the slow wave of the EEG (Czigler et al., 2008; Engedal et al., 2020; Han & Youn, 2022; Jeong, Youn, Sung, & Kim, 2021; Ya et al., 2015). Spectral analysis of the EEG is used to develop simple diagnostic algorithms that can identify MCI. However, finding distinct EEG signals with strong neuropathology and neuropsychology correlations are necessary for reliable and accurate assessments.

In a broad frequency range, spectral- and EEG-complexity alterations can be detected even in the early stages of AD. However, applying traditional EEG analysis techniques along with QEEG may further increase the likelihood of making an early diagnosis of AD (Czigler et al., 2008; Ya et al., 2015). A recent study by Engedal et al. (2020) concluded that applying QEEG using the statistical pattern recognition (SPR) method could predict conversion to dementia in patients with SCD and MCI. The authors also suggested adding this method to other routine diagnostic methods like cognitive tests and other biomarkers to benefit from the discriminant power of the proposed method. A study by Jeong et al. (2021) demonstrated that SCD participants had more frontal delta waves and less occipital alpha1 compared to control participants. SCD is one of AD's early clinical signs and is linked to nerve degeneration. Hamilton et al. (2021) demonstrated that QEEG slowing may precede the onset of dementia in prospectively identified MCI. In addition, another study by Han and Youn (2022) used QEEG for monitoring treatment response in MCI patients. Future studies should utilise QEEG characteristics, neuroimaging, and/or machine learning to better collect and show neural network activation.

**EEG-based machine learning and mild cognitive impairment**

Machine learning is a set of algorithms that enables us to automatically detect patterns in the data and predict the results of measurements (Bishop, 2006; Hastie, Tibshirani, & Friedman, 2009; Murphy, 2013). It is frequently used in neural engineering and neuroscience experiments to: (1) compare different circumstances (Lemm, Schafer, & Curio, 2004; Müller et al., 2008), (2) make a diagnosis of a disease (Dauwan et al., 2018; Kumar, Dewal, & Anand, 2014), or (3) find electrophysiological alterations connected to behaviour (Buriro, Shoorangiz, Weddell, & Jones, 2018; Davidson, Jones, & Peiris, 2007). Machine learning algorithms have been developed to extract information from the EEG in order to identify different brain states and aid in the diagnosis of several conditions (such as epilepsy, AD, and schizophrenia) (Shoorangiz, Weddell, & Jones, 2021).

Several studies have described EEG-based machine learning frameworks for the diagnosis of AD, MCI, and other types of dementia. Mcbride et al. (2014) established an EEG-based classification system using spectrum and complexity analysis. The complexity analysis includes computational activity, mobility, complexity, sample entropy, and lempel-Ziv characteristics, while the spectral analysis involves extracting features from delta, theta, alpha1, alpha2, gamma, beta1, and beta2 subbands. The support vector machine (SVM) classifier was found to have an average accuracy of 79.2% for MCI classification. Kashefpoor, Rabbani, and Barekatain (2016) extracted EEG spectral parameters using delta, theta, alpha1, alpha2, gamma, beta1, and beta2 subbands, as well as 19 spectral characteristics from a 19-channel EEG data. A correlation-based algorithm identified the best discriminative characteristics. The best classification accuracy was 88.89% utilising neuro-fuzzy (NF) and k-nearest neighbour (KNN) classifiers, (Kashefpoor et al., 2016). Kashefpoor, Rabbani, and Barekatain (2019), in a different study, employed a 19-channel EEG data and supervised dictionary learning techniques like label-consistent K-SVD (LC-KSVD) and correlation-based label-consistent K-SVD (CLC-KSVD) to diagnose MCI and achieved the best accuracy at 88.9%.

Khatun, Morshed, and Bidelman (2019) developed a single-channel EEG method for diagnosing MCI using speech-evoked brain responses. The top 25 characteristics were picked using the random forest (RF) method after temporal and spectral domain analysis extracted 590 from recorded sounds. The best classification result they found was 87.9% accuracy using logistic regression (LR) and SVM models. Yin, Cao, Siuly, and Wang (2019) also suggested a spectral-temporal analysis-based MCI diagnosis technique. This method employs spectral-temporal analysis to extract features and the 3-D evaluation algorithm to produce an optimum feature subset. In the study, the SVM classifier had 96.94% accuracy (Yin et al. 2019).

Sharma, Kolekar, Jha, & Kumar (2019) used power spectral density (PSD), skewness, kurtosis, spectral skewness, spectral kurtosis, spectral crest factor, spectral entropy (SE), and fractal dimension (FD) to develop an EEG-based MCI detection method. Open eyes, closed eyes, finger tapping test (FTT), and continuous performance test were the conditions used to collect the dataset. SVM's classification accuracy was 96.94% (Sharma et al., 2019). Later, Siuly et al. (2020) established an MCI patient-identification system using extreme learning machine (ELM), SVM, and KNN classification models. The ELM classifier achieved the best classification result, with 98.78% accuracy (Siuly et al., 2020).

The spectral-power-based task-induced intra-subject EEG variability extracted by the sequential forward selection (SFS) algorithm-based feature selection and SVM classification has been shown to have the potential to serve as a neurophysiological feature for the early detection of MCI in individuals (Trinh et al., 2021). Rossini, Miraglia, and Vecchio (2022), on the other hand, used a combination of graph analysis tools and SVM classification to identify the distinctive features of physiological/pathological brain ageing focusing on functional connectivity networks evaluated on EEG data and other biomarkers.

Hsiao et al. (2021) proposed a novel EEG feature, the kernel eigen-relative-power (KERP) feature, for achieving high classification accuracy of MCI (90.2%) when used in combination with an SVM classifier. Meanwhile, Oltu, Akşahin, and Kibaroğlu (2021) proposed an EEG-based categorization for MCI, AD, and healthy persons using discrete wavelet transform (DWT), PSD, and interhemispheric coherence for feature extraction, and Bagged Trees for classification. The study's best result was 96% (Oltu et al. 2021). Movahed and Rezaeian (2022) proposed an automatic diagnosis of MCI using 0-spectral, functional connectivity, and nonlinear EEG-based features. Ten-fold cross-validation was used to assess metrics including accuracy (AC), sensitivity (SE), specificity (SP), F1-score (F1), and false discovery rate (FDR). The best performance of the proposed framework employing the linear support vector machine (LSVM) classifier and the combination of all feature sets have been reported, with an average AC of 99.4%, SE of 98.8%, SP of 100%, F1 of 99.4%, and FDR of 0%. Based on the achieved results, it is possible to develop and use a computer-aided diagnostic (CAD) tool for clinical purposes.

## Conclusion

This scoping review examined 2310 articles on EEG and MCI. The articles were divided into four broad themes, including ERP/EEG, epilepsy, QEEG, and EEG-based machine learning. According to numerous studies the ERP/EEG, QEEG and EEG-based machine learning framework can be widely used for high-accuracy seizure and MCI detector.